# DYNAMIC CAUSAL MODELLING OF COVID-19

Karl J. Friston[1], Thomas Parr[1], Peter Zeidman[1], Adeel Razi[2], Guillaume Flandin[1], Jean Daunizeau[3], Oliver J. Hulme[4,5], Alexander J. Billig[6], Vladimir Litvak[1], Rosalyn J. Moran[7], Cathy J. Price[1] and Christian Lambert[1]

[1]*The Wellcome Centre for Human Neuroimaging, University College London, UK*
[2]*Turner Institute for Brain and Mental Health & Monash Biomedical Imaging, Monash University, Clayton, Australia*
[3]*Institut du Cerveau et de la Moelle épinière, INSERM UMRS 1127, Paris, France*
[4]*Danish Research Centre for Magnetic Resonance, Centre for Functional and Diagnostic Imaging and Research, Copenhagen University Hospital Hvidovre, Kettegaard Allé 30, Hvidovre, Denmark.*
[5]*London Mathematical Laboratory, 8 Margravine Gardens, Hammersmith, UK*
[6]*Ear Institute, University College London, UK*
[7]*Centre for Neuroimaging Science, Department of Neuroimaging, IoPPN, King's College London, UK*

***E-mails***: *Karl Friston, k.friston@ucl.ac.uk; Thomas Parr, thomas.parr.12@ucl.ac.uk; Peter Zeidman, peter.zeidman@ucl.ac.uk; Adeel Razi, adeel.razi@monash.edu; Guillaume Flandin, g.flandin@ucl.ac.uk; Jean Daunizeau, jean.daunizeau@googlemail.com; Ollie Hulme, oliverh@drcmr.dk; Alexander Billig, a.billig@ucl.ac.uk; Vladimir Litvak, v.litvak@ucl.ac.uk; Rosalyn Moran, rosalyn.moran@kcl.ac.uk; Cathy Price, c.j.price@ucl.ac.uk; Christian Lambert, christian.lambert@ucl.ac.uk*

## Abstract

This technical report describes a dynamic causal model of the spread of coronavirus through a population. The model is based upon ensemble or population dynamics that generate outcomes, like new cases and deaths over time. The purpose of this model is to quantify the uncertainty that attends predictions of relevant outcomes. By assuming suitable conditional dependencies, one can model the effects of interventions (e.g., social distancing) and differences among populations (e.g., herd immunity) to predict what might happen in different circumstances. Technically, this model leverages state-of-the-art variational (Bayesian) model inversion and comparison procedures, originally developed to characterise the responses of neuronal ensembles to perturbations. Here, this modelling is applied to epidemiological populations—to illustrate the kind of inferences that are supported and how the model *per se* can be optimised given timeseries data. Although the purpose of this paper is to describe a modelling protocol, the results illustrate some interesting perspectives on the current pandemic; for example, the nonlinear effects of herd immunity that speak to a self-organised mitigation process.

# Contents

# Introduction

The purpose of this paper is to show how dynamic causal modelling can be used to make predictions—and test hypotheses—about the ongoing coronavirus pandemic (Huang et al., 2020; Wu et al., 2020; Zhu et al., 2020). It should be read as a technical report[1], written for people who want to understand what this kind of modelling has to offer (or just build an intuition about modelling pandemics). It contains a sufficient level of technical detail to implement the model using Matlab (or its open source version Octave), while explaining things heuristically for non-technical readers. The examples in this report are used to showcase the procedures and subsequent inferences that can be drawn. Having said this, there are some quantitative results that will be of general interest. These results are entirely conditional upon the model used.

Dynamic causal modelling (DCM) refers to the characterisation of coupled dynamical systems in terms of how observable data are generated by unobserved (i.e., latent or hidden) causes (Friston et al., 2003; Moran et al., 2013). Dynamic causal modelling subsumes state estimation and system identification under one Bayesian procedure, to provide probability densities over unknown latent states (i.e., state estimation) and model parameters (i.e., system identification), respectively. Its focus is on estimating the uncertainty about these estimates to quantify the evidence for competing models, and the confidence in various predictions. In this sense, DCM combines data assimilation and uncertainty quantification within the same optimisation process. Specifically, the posterior densities (i.e., Bayesian beliefs) over states and parameters—and the precision of random fluctuations—are optimised by maximising a variational bound on the model's marginal likelihood, also known as *model evidence*. This bound is known as variational free energy or the evidence lower bound (ELBO) in machine learning (Friston et al., 2007; Hinton and Zemel, 1993; MacKay, 1995; Winn and Bishop, 2005).

---

[1] Prepared as a proof of concept for submission to the SPI-M ([https://www.gov.uk/government/groups/scientific-pandemic-influenza-subgroup-on-modelling](https://www.gov.uk/government/groups/scientific-pandemic-influenza-subgroup-on-modelling)) and the RAMP (Rapid Assistance in Modelling the Pandemic) initiative ([https://royalsociety.org/topics-policy/Health-and-wellbeing/ramp/](https://royalsociety.org/topics-policy/Health-and-wellbeing/ramp/)).

Intuitively, this means one is trying to optimise probabilistic beliefs—about the unknown quantities generating some data—such that the (marginal) likelihood of those data is as large as possible. The marginal likelihood[2] or model evidence can always be expressed as *accuracy* minus *complexity*. This means that the best models provide an accurate account of some data as *simply as possible*. Therefore, the model with the highest evidence is not necessarily a description of the process generating data: rather, it is the simplest description that provides an accurate account of those data. In short, it is 'as if' the data were generated by this kind of model. Importantly, models with the highest evidence will generalise to new data and preclude overfitting, or overconfident predictions about outcomes that have yet to be measured. In light of this, it is imperative to select the parameters or models that maximise model evidence or variational free energy (as opposed to goodness of fit or accuracy). However, this requires the estimation of the uncertainty about model parameters and states, which is necessary to evaluate the (marginal) likelihood of the data at hand. This is why estimating uncertainty is crucial. Being able to score a model—in terms of its evidence—means that one can compare different models of the same data. This is known as *Bayesian model comparison* and plays an important role when testing different models or hypotheses about how the data are caused. We will see examples of this later. This aspect of dynamic causal modelling means that one does not have to commit to a particular form (i.e., parameterisation) of a model. Rather, one can explore a repertoire of plausible models and let the data decide which is the most apt.

Dynamic causal models are *generative models* that generate consequences (i.e., data) from causes (i.e., hidden states and parameters). The form of these models can vary depending upon the kind of system at hand. Here, we use a ubiquitous form of model; namely, a mean field approximation to loosely coupled ensembles or populations. In the neurosciences, this kind of model is applied to populations of neurons that respond to experimental stimulation (Marreiros et al., 2009; Moran et al., 2013). Here, we use the same mathematical approach to model a population of individuals and their response to an epidemic. The key idea behind these (mean field) models is that the constituents of the ensemble are exchangeable; in the sense that sampling people from the population at random will give the same average as following one person over a long period of time. Under this assumption[3], one can then work out, analytically, how the probability distribution over various states of people evolve over time, e.g., whether someone was infected or not. This involves parameterising the probability that people will transition from one state to another. By assuming the population is large, one can work out the likelihood of observing a certain number of people who were infected, given the probabilistic state of the population at that point in time. In turn, one can work out the probability of a sequence or timeseries of new cases. This is the kind of generative model used here, where the latent states were chosen to generate the data that are—or could be—used to track a pandemic. Figure 1 provides an overview of this model. In terms of epidemiological models, this can be regarded as an extended SEIR (susceptible, exposed, infected and recovered) compartmental model (Kermack et al., 1997). Please see (Kucharski et al., 2020) for an application of this kind of model to COVID-19[4].

---

[2] The marginal likelihood is the likelihood having marginalised (i.e., averaged) over unknown quantities like states and parameters: i.e., the probability of having observed some data under a particular model.

There are number of advantages to using a model of this sort. First, it means that one can include every variable that 'matters', such that one is not just modelling the spread of an infection but an ensemble response in terms of behaviour (e.g., social distancing). This means that one can test hypotheses about the contribution of various responses that are installed in the model—or what would happen under a different kind of response. A second advantage of having a generative model is that one can evaluate its evidence in relation to alternative models, and therefore optimise the structure of the model itself. For example, does social distancing behaviour depend upon the number of people who are infected? Or, does it depend on how many people have tested positive for COVID-19? (this question is addressed below). A third advantage is more practical, in terms of data analysis: because we are dealing with ensemble dynamics, there is no need to create multiple realisations or random samples to estimate uncertainty. This is because the latent states are not the states of an individual but the sufficient statistics of a probability distribution over individual states. In other words, we replace random fluctuations in hidden states with hidden states that parameterise random fluctuations. The practical consequence of this is that one can fit these models quickly and efficiently—and perform model comparisons over thousands of models. A fourth advantage is that, given a set of transition probabilities, the ensemble dynamics are specified completely. This has the simple but important consequence that the only unknowns in the model are the parameters of these transition probabilities. Crucially, in this model, these do not change with time. This means that we can convert what would have been a very complicated, nonlinear state space model for data assimilation into a nonlinear mapping from some unknown (probability transition) parameters to a sequence of observations. We can therefore make precise predictions about the long-term future, under particular circumstances. This follows because the only uncertainty about outcomes inherits from the uncertainty about the parameters, which do not change with time. These points may sound subtle; however, the worked examples below have been chosen to illustrate these properties.

---

[3] Technically, this property reflects ergodicity that is a consequence of a weakly mixing system: Birkhoff, G.D., 1931. Proof of the ergodic theorem. Proc Natl Acad Sci USA 17, 656-660. Having said this, this model aims to make ensemble level predictions. Because ergodicity may not necessarily hold in reality, the ensemble level projections should not be interpreted as predictions for individual experiences of the epidemic

[4] Conventional (e.g., SEIR) compartmental models in epidemiology usually consider a single attribute of the population (e.g., infection status), such that the distribution over all states sums to one. In contrast, the DCM used in this work considers multiple attributes, where both the joint and marginal distributions (over each factor) sum to one.

This technical report comprises four sections. The first details the generative model, with a focus on the conditional dependencies that underwrite the ensemble dynamics generating outcomes. The outcomes in question here pertain to a regional outbreak. This can be regarded as a generative model for the first wave of an epidemic in a large city or metropolis. This section considers variational model inversion and comparison, under hierarchical models. In other words, it considers the distinction between (first level) models of an outbreak in one country and (second level) models of differences among countries, in terms of model parameters. The second section briefly surveys the results of second level (between-country) modelling, looking at those aspects of the model that are conserved over countries (i.e., random effects) and those which are not (i.e., fixed effects). The third section then moves on to the dynamics and predictions for a single country; here, the United Kingdom. It considers the likely outcomes over the next few weeks and how confident one can be about these outcomes, given data from all countries to date. This section drills down on the parameters that matter in terms of affecting death rates. It presents a sensitivity analysis that establishes the contribution of parameters or causes in the model to eventual outcomes. It concludes by looking at the effects of social distancing and herd immunity. The final section concludes with a consideration of predictive validity by comparing predicted and actual outcomes.

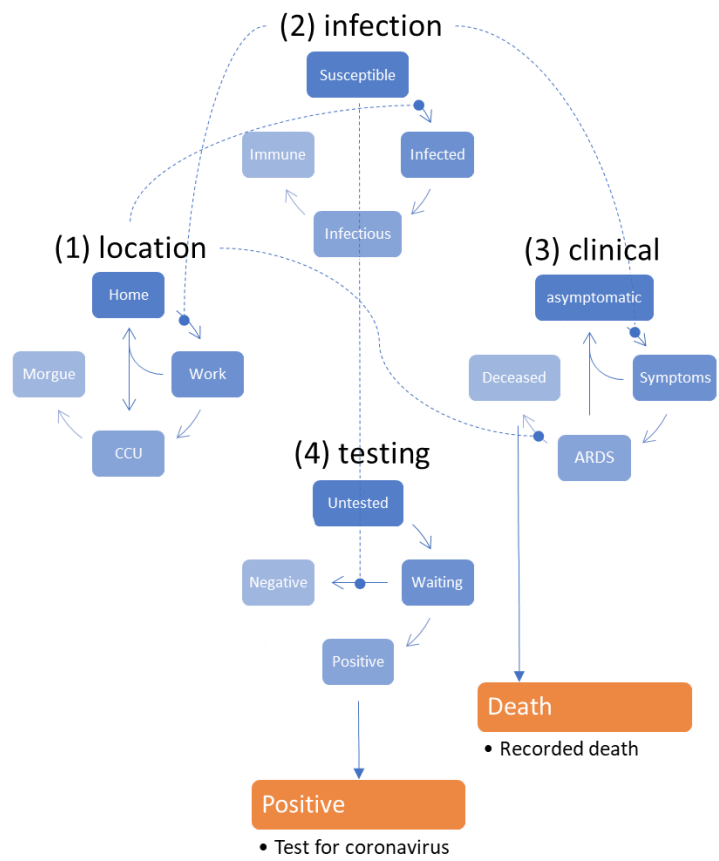

**FIGURE 1**

**Figure 1: generative model.** This figure is a schematic description of the generative model used in subsequent analyses. In brief, this compartmental model generates timeseries data based on a mean field approximation to ensemble or population dynamics. The implicit probability distributions are over four latent factors, each with four levels or states. These factors are sufficient to generate measurable outcomes; for example, the number of new cases or the proportion of people infected. The first factor is the location of an individual, who can be at *home*, at *work*, in a critical care unit (*CCU*) or in the *morgue*. The second factor is infection status; namely, *susceptible* to infection, *infected*, *infectious* or *immune*. This model assumes that there is a progression from a state of susceptibility to immunity, through a period of (pre-contagious) infection to an infectious (contagious) status. The third factor is clinical status; namely, *asymptomatic*, *symptomatic*, *acute respiratory distress syndrome (ARDS)* or *deceased*. Again, there is an assumed progression from asymptomatic to ARDS, where people with ARDS can either recover to an asymptomatic state or not. Finally, the fourth factor represents diagnostic or testing status. An individual can be *untested* or *waiting* for the results of a test that can either be *positive* or *negative*. With this setup, one can be in one of four places, with any infectious status, expressing symptoms or not, and having test results or not. Note that—in this construction—it is possible to be infected and yet be asymptomatic. However, the marginal distributions are not independent, by virtue of the dynamics that describe the transition among states within each factor. Crucially, the transitions within any factor depend upon the marginal distribution of other factors. For example, the probability of becoming infected, given that one is susceptible to infection, depends upon whether one is at home or at work. Similarly, the probability of developing symptoms depends upon whether one is infected or not. The probability of testing negative depends upon whether one is susceptible (or immune) to infection, and so on. Finally, to complete the circular dependency, the probability of leaving home to go to work depends upon the number of infected people in the population, mediated by social distancing. The curvilinear arrows denote a conditioning of transition probabilities on the marginal distributions over other factors. These conditional dependencies constitute the mean field approximation and enable the dynamics to be solved or integrated over time. At any point in time, the probability of being in any combination of the four states determines what would be observed at the population level. For example, the occupancy of the deceased level of the clinical factor determines the current number of people who have recorded deaths. Similarly, the occupancy of the positive level of the testing factor determines the expected number of positive cases reported. From these expectations, the expected number of new cases per day can be generated. A more detailed description of the generative model—in terms of transition probabilities—can be found in in the main text.

# The generative model

This section describes the generative model summarised schematically in Figure 1, while the data used to invert or fit this model are summarised in Figure 2. These data comprise global (worldwide) timeseries from countries and regions from the initial reports of positive cases in China to the current day[5].

---

[5] These data are available from: https://github.com/CSSEGISandData/COVID-19.

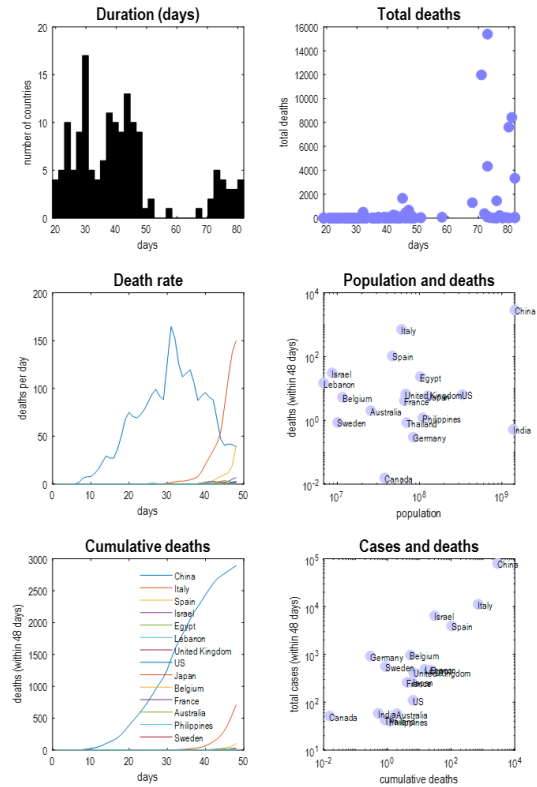

The generative model is a mean field model of ensemble dynamics. In other words, it is a state space model where the states correspond to the sufficient statistics (i.e., parameters) of a probability distribution over the states of an ensemble or population—here, a population of people who are in mutual contact at some point in their daily lives. This kind of model is used routinely to model populations of neurons, where the ensemble dynamics are cast as density dynamics, under Gaussian assumptions about the probability densities; e.g., (Marreiros et al., 2009). In other words, a model of how the mean and covariance of a population affects itself and the means and covariances of other populations. Here, we will focus on a single population and, crucially, use a discrete state space model. This means that we will be dealing with the sufficient statistics (i.e. expectations) of the probability of being in a particular state at any one time. This renders the model a compartmental model (Kermack et al., 1997), where each state corresponds to a compartment. These latent states evolve according to transition probabilities that embody the causal influences and conditional dependencies that lend an epidemic its characteristic form. Our objective is to identify the right conditional dependencies—and form posterior beliefs about the model parameters that mediate these dependencies. Having done this, we can then simulate an entire trajectory into the distant future, even if we are only given data about the beginning of an outbreak[6].

The model considers four different sorts of states (i.e., factors) that provide a description of any individual— sampled at random—that is sufficient to generate the data at hand. In brief, these factors were chosen to be as conditionally independent as possible to ensure an efficient estimation of the model parameters[7]. The four factors were an individual's *location*, *infection* status, *clinical* status and *diagnostic* status. In other words, we considered that any member of the population can be characterised in terms of where they were, whether they were infected, infectious or immune, whether they were showing mild and severe or fatal symptoms, and whether they had been tested with an ensuing positive or negative result. Each of these factors had four levels. For example, the location factor was divided into *home*, *work*, *critical care unit*, and the *morgue*. These states should not be taken too literally. For example, *home* stands in for anywhere that has a limited risk of exposure to, or contact with, an infected person (e.g., in the domestic home, in a non-critical hospital bed, in a care home, *etc*). *Work* stands in for anywhere that has a larger risk of exposure to—or contact with—an infected person and therefore covers non-work activities, such as going to the supermarket or participating in team sports. Similarly, designating someone as severely ill with acute respiratory distress syndrome (ARDS) is meant to cover any life-threatening conditions that would invite admission to intensive care.

Having established the state space, we can now turn to the causal aspect of the dynamic causal model. The causal structure of these models depends upon the dynamics or transitions from one state or another. It is at this point that a mean field approximation can be used. Mean field approximations are used widely in physics to approximate a full (joint) probability density with the product of a series of marginal densities (Bressloff and Newby, 2013; Marreiros et al., 2009; Schumacher et al., 2015; Zhang et al., 2018). In this case, the factorisation is fairly subtle: we will factorise the transition probabilities, such that the probability of moving among states—within each factor—depends upon the marginal distribution of other factors (with one exception). For example, the probability of developing *symptoms* when *asymptomatic* depends on, and only on, the probability that I am *infected*. In what follows, we will step through the conditional probabilities for each factor to show how the model is put together (and could be changed).

---

[6] Note that part of the uncertainty about latent states inherits from uncertainty about how outcomes are generated; for example, uncertainty about whether reported death rates are a true reflection of actual death rates.

[7] This involves examining the eigenvectors of the posterior correlation matrix, to preclude marked posterior correlations.

## Transition probabilities and priors

The first factor has four levels, *home*, *work*, *CCU* and the *morgue*. People can leave home but will always return (with unit probability) over a day. The probability of leaving home has a (prior) baseline rate of one third but is nuanced by any social distancing imperatives. These imperatives are predicated on the proportion of the population that is currently infected, such that the social distancing parameter (an exponent) determines the probability of leaving home[8]. For example, social distancing is modelled as the propensity to leave home and expose oneself to interpersonal contacts. This can be modelled with the following transition probability:

$$P(loc_{t+1} = work \mid loc_t = home, clin_t = asymptomatic_t) = \theta_{out}(1 - p_{infected}^{inf})^{\theta_{sde}} \tag{1.1}$$

This means that the probability of leaving home, given I have no symptoms, is the probability I would have gone out normally, multiplied by a decreasing function of the proportion of people in the population who are infected. Formally, this proportion is the marginal probability of being infected, where the marginal probability of a factor is an average over the remaining factors. The marginal probability $p^\ell$ of the *location* factor is as follows:

$$p_{\ell ijk} = P(loc = \ell, inf = i, clin = j, test = k)$$
$$p^\ell = \sum_{ijk} p_{\ell ijk}$$
$$\sum_u p_u^\ell = 1$$

$$p^{loc} = [p_{home}^{loc}, p_{work}^{loc}, p_{CCU}^{loc}, p_{morgue}^{loc}]$$
$$p^{inf} = [p_{susceptible}^{inf}, p_{infected}^{inf}, p_{infectious}^{inf}, p_{immune}^{inf}]$$
$$p^{clin} = [p_{asymptomatic}^{clin}, p_{symptomatic}^{clin}, p_{ADRS}^{clin}, p_{deceased}^{clin}]$$
$$p^{test} = [p_{untested}^{test}, p_{waiting}^{test}, p_{positive}^{test}, p_{negative}^{test}]$$

$$\tag{1.2}$$

Where the final four equalities define each factor or state in the model. The parameters in this social

---

distancing model are the probability of leaving home every day $(\theta_{out})$ and the social distancing exponent $(\theta_{sde})$.

The only other two places one can be are in a *CCU* or the *morgue*. The probability of moving to critical care depends upon bed (i.e., hospital) availability, which is modelled as a sigmoid function of the occupancy of this state (i.e., the probability that a *CCU* bed is occupied) and a bed capacity parameter (a threshold). If one has severe symptoms, then one stays in the *CCU*. Finally, the probability of moving to the morgue depends on, and only on, being *deceased*. Note that all these dependencies are different states of the *clinical* factor (see below). This means we can write the transition probabilities among the *location* factor for each level of the *clinical* factor as follows (with a slight abuse of notation):

$$P = \theta_{out}(1 - p_{infected}^{inf})^{\theta_{sde}}$$
$$Q = \sigma(p_{CCU}^{loc}, \theta_{cap})$$

$$P\left(loc_{t+1} \mid loc_t, clin_t = asymptomatic\right) = \begin{bmatrix} 1-P & 1 & 1 & \\ P & & & \\ & & & \\ & & & 1 \end{bmatrix}$$

$$P\left(loc_{t+1} \mid loc_t, clin_t = symptomatic\right) = \begin{bmatrix} 1 & 1 & 1 & \\ & & & \\ & & & \\ & & & 1 \end{bmatrix}$$

$$P\left(loc_{t+1} \mid loc_t, clin_t = ARDS\right) = \begin{bmatrix} 1-Q & 1-Q & & \\ & & & \\ Q & Q & 1 & \\ & & & 1 \end{bmatrix} \qquad (1.3)$$

$$P\left(loc_{t+1} \mid loc_t, clin_t = deceased\right) = \begin{bmatrix} & & & \\ & & & \\ & & & \\ 1 & 1 & 1 & 1 \end{bmatrix}$$

Here, the columns and rows of each transition probability matrix are ordered: home, work, CCU, morgue. The column indicates the current location and the row indicates the next location. Parameter $\theta_{cap}$ is bed capacity threshold and $\sigma(s, \theta) = (1 + e^{4(\frac{s}{\theta}-1)})^{-1}$ is a decreasing sigmoid function. In brief, these transition probabilities mean that I will go out when *asymptomatic*, unless social distancing is in play. However, when I have symptoms I will stay at home, unless I am hospitalised with acute respiratory distress. I remain in critical care unless I recover and go home or die and move to the morgue, where I stay. Technically, the *morgue* is an absorbing state.

In a similar way, we can express the probability of moving between different states of infection (i.e., *susceptible*, *infected*, *infectious* and *immune*) as follows:

$$P = (1 - \theta_{trn} \cdot p_{infectious}^{inf})$$

$$P\left(inf_{t+1} \mid inf_t, loc_t = home\right) = \begin{bmatrix} P^{\theta_{Rin}} & & & \\ 1 - P^{\theta_{Rin}} & \theta_{inf} & & \\ & 1 - \theta_{inf} & \theta_{con} & \\ & & 1 - \theta_{con} & 1 \end{bmatrix}$$

$$P\left(inf_{t+1} \mid inf_t, loc_t = work\right) = \begin{bmatrix} P^{\theta_{Rou}} & & & \\ 1 - P^{\theta_{Rou}} & \theta_{inf} & & \\ & 1 - \theta_{inf} & \theta_{con} & \\ & & 1 - \theta_{con} & 1 \end{bmatrix}$$

$$P\left(inf_{t+1} \mid inf_t, loc_t = CCU\right) = \begin{bmatrix} 1 & & & \\ & \theta_{inf} & & \\ & 1 - \theta_{inf} & \theta_{con} & \\ & & 1 - \theta_{con} & 1 \end{bmatrix}$$

$$P\left(inf_{t+1} \mid inf_t, loc_t = morgue\right) = \begin{bmatrix} & & & \\ & & & \\ & & & \\ 1 & 1 & 1 & 1 \end{bmatrix}$$

(1.4)

These transition probabilities mean that when susceptible, the probability of becoming infected depends upon the number of social contacts—which depends upon the proportion of time spent at home. This dependency is parameterised in terms of a transition probability per contact ($\theta_{trn}$) and the expected number of contacts at home ($\theta_{Rin}$) and work ($\theta_{Rou}$)[9]. Once infected, one remains in this state for a period of time that is parameterised by a transition rate ($\theta_{inf}$). This parameterisation illustrates a generic property of transition probabilities; namely, an interpretation in terms of rate constants and, implicitly, time constants. The rate parameter $\theta$ is related to the rate constant $\kappa$ and time constant $\tau$ according to:

$$\theta = \exp(-\kappa) = \exp(-\tfrac{1}{\tau}) \leq 1 : \forall \tau > 0$$

(1.5)

In other words, the probability of staying in any one state is determined by the characteristic length of time that state is occupied. This means that the rate parameter above can be specified, *a priori*, in terms of the number of days we expect people to be infected, before becoming infectious. Similarly, we can parameterise the transition from being infectious to being immune in terms of a typical period of being contagious, assuming that immunity is enduring and precludes reinfection[10]. Note that in the model, everybody in the morgue is treated as having has acquired immunity. The transitions among clinical states depend upon both the infection status and location as follows:

$$\theta_{fat} = \begin{cases} \theta_{fat} & loc = CCU \\ 1 - \theta_{sur} & loc \neq CCU \end{cases}$$

$$P\left(clin_{t+1} \mid clin_t, inf_t \in \{susceptible, immune\}\right) = \begin{bmatrix} 1 & 1 & 1 & \\ & & & \\ & & & \\ & & 1 & \end{bmatrix} \quad (1.6)$$

$$P\left(clin_{t+1} \mid clin_t, inf_t \in \{infected, infectious\}\right) = \begin{bmatrix} 1-\theta_{dev} & (1-\theta_{sym})(1-\theta_{sev}) & (1-\theta_{rds})(1-\theta_{fat}) & \\ \theta_{dev} & \theta_{sym} & & \\ & (1-\theta_{sym})\theta_{sev} & \theta_{rds} & \\ & & (1-\theta_{rds})\theta_{fat} & 1 \end{bmatrix}$$

---

[9] Here, $P = (1 - \theta_{trn} \cdot p_{infectious}^{inf})$ can be interpreted as a probability of eluding infection with each interpersonal contact, such that the probability of remaining uninfected after $\theta_R$ contacts is given by $P^{\theta_R}$. Note, that there is no distinction between people at home and at work; both are equally likely to be infectious.

The transitions among clinical states (i.e., *asymptomatic*, *symptomatic*, *ARDS* and *deceased*) are relatively straightforward: if I am not infected (i.e., *susceptible* or *immune*) I will move to the asymptomatic state, unless I am dead. However, if I am infected (i.e., *infected* or *infectious*), I will develop symptoms with a particular probability ($\theta_{dev}$). Once I have developed symptoms, I will remain symptomatic and either recover to an asymptomatic state or develop acute respiratory distress with a particular probability ($\theta_{sev}$). The parameterisation of these transitions depends upon the typical length of time that I remain symptomatic ($\theta_{sym}$); similarly, when in acute respiratory distress ($\theta_{rds}$). However, I may die following ARDS, with a probability that depends upon whether I am in a CCU, or elsewhere. This is the exception (mentioned above) to the conditional dependencies on marginal densities. Here, the probability of dying ($\theta_{fat}$) depends on being infected and my location: I am more likely to die of ARDS, if I am not in CCU, where $\theta_{sur}$ is the probability of surviving at home. The implication here is that the transition probabilities depend upon two marginal densities, as opposed to one for all the other factors: see the first equality in (1.6). Please refer to Table 1 for details of the model parameters.

Finally, we turn to diagnostic testing status (i.e., *untested*, *waiting* or *positive* versus *negative*). The transition probabilities here are parameterised in terms of test availability ($\theta_{tft}$, $\theta_{sen}$). and the probability that I would have been tested anyway, which is relatively smaller, if I am asymptomatic ($\theta_{tes}$). Test availability is a decreasing sigmoid function of the number of people who are waiting (with a delay $\theta_{del}$) for their results.

---

[10] Although cases of double positive diagnoses have reported, these may reflect false test results rather than re-infection *per se*. Having said this, it would be straightforward to include a transition from immunity to susceptible, with a suitably small transition rate to model the decay of immunity, or viral mutation. Similarly, one can incorporate transition from asymptomatic to symptomatic, when immune, to model partial resistance.

I can only move from being *untested* to *waiting*. After this, I can only go into *positive* or *negative* test states, depending upon whether I have the virus (i.e., *infected* or *infectious*) or not[11].

$$P = \theta_{sen}\sigma(p^{test}_{waiting}, \theta_{tft})$$

$$P\left(test_{t+1} \mid test_t, inf_t \in \{susceptible, immune\}\right) = \begin{bmatrix} 1-\theta_{tes}P & & & \\ \theta_{tes}P & \theta_{del} & & \\ & & 1 & \\ & 1-\theta_{del} & 1 & \end{bmatrix}$$

$$P\left(test_{t+1} \mid test_t, inf_t \in \{infected, infectious\}\right) = \begin{bmatrix} 1-P & & & \\ P & \theta_{del} & & \\ & 1-\theta_{del} & 1 & \\ & & & 1 \end{bmatrix}$$

(1.7)

We can now assemble these transition probabilities into a probability transition matrix, and iterate from the first day to some time horizon, to generate a sequence of probability distributions over the joint space of all factors:

$$p_{t+1} = T(\theta, p_t)p_t \tag{1.8}$$

Notice that this is a completely deterministic state space model, because all the randomness is contained in the probabilities. Notice also that the transition probability matrix $T$ is both state *and time* dependent, because the transition probabilities above depend on marginal probabilities. Technically, (1.8) is known as a *master equation* (Seifert, 2012; Vespignani and Zapperi, 1998; Wang, 2009) and forms the basis of the dynamic part of the dynamic causal model.

This model of transmission supports an *effective reproduction number or rate, R,* which summarises how many people I am likely to infect, if I am infected. This depends upon the probability that any contact will cause an infection, the probability that the contact is susceptible to infection and number of people I contact:

$$R = \theta_{trn} \cdot p^{inf}_{susceptible} \cdot (p^{loc|infectious}_{home}\theta_{Rin} + p^{loc|infectious}_{work}\theta_{Rou}) \cdot \tau_{con} \tag{1.9}$$

---

[11] Notice that this model is configured for new cases that are reported based on buccal swabs (i.e., am I currently infected?), not tests for antibody or immunological status. A different model would be required for forthcoming tests of immunity (i.e., have I been infected?). Furthermore, one might consider the sensitivity and specificity of any test by including sensitivity and specificity in (1.7). For example, 1 in 3 tests may be false negatives; especially, when avoiding bronchoalveolar lavage to minimise risk to clinicians: Wang, W., Xu, Y., Gao, R., Lu, R., Han, K., Wu, G., Tan, W., 2020b. Detection of SARS-CoV-2 in Different Types of Clinical Specimens. JAMA.

In this approximation, the number of contacts I make is a weighted average of the number of people I could infect at home and the number of people I meet outside, per day, times the number of days I am contagious. The effective reproduction rate is not a biological rate constant. However, it is a useful epidemiological summary statistic that indicates how quickly the disease spreads through a population. When less than one, the infection will decay to an endemic equilibrium. We will use this measure later to understand the role of herd immunity.

This completes the specification of the generative model of latent states. A list of the parameters and their prior means (and variances) is provided in Table 1. Notice that all of the parameters are scale parameters, i.e., they are rates or probabilities that cannot be negative. To enforce these positivity constraints, one applies a log transform to the parameters during model inversion or fitting. This has the advantage of being able to simplify the numerics using Gaussian assumptions about the prior density (via a lognormal assumption). In other words, although the scale parameters are implemented as probabilities or rates, they are estimated as log parameters, denoted by $\vartheta = \ln\theta$. Note that prior variances are specified for log parameters. For example, a variance of 1/64 corresponds to a prior confidence interval of ~25% and can be considered weakly informative.

### Table 1

Parameters of the COVID-19 model and priors, $N(\eta, C)$

(NB: prior means are for scale parameters $\theta = \exp(\vartheta)$ )

| Number | Parameter | Mean | Variance | Description |
|--------|-----------|------|----------|-------------|
| 1 | $\theta_n$ | 1 | 1/4 | Number of initial cases |
| 2 | $\theta_N$ | 1 | 1/16 | Effective population size (millions) |
| 3 | $\theta_m$ | $10^{-6}$ | 0 | Herd immunity (proportion) |
| **Location** | | | | |
| 4 | $\theta_{out}$ | 1/3 | 1/64 | Pr(*work* \| *home*): probability of going out |
| 5 | $\theta_{sde}$ | 32 | 1/64 | Social distancing exponent |
| 6 | $\theta_{cap}$ | 128/100000 | 1/64 | CCU capacity threshold (per capita) |
| **Infection** | | | | |
| 7 | $\theta_{Rin}$ | 3 | 1/64 | Effective number of contacts: home |
| 8 | $\theta_{Rou}$ | 48 | 1/64 | Effective number of contacts: work |
| 9 | $\theta_{trn}$ | 1/4 | 1/64 | Pr(*contagion* \| *contact*) |
| 10 | $\theta_{inf} = \exp(-\frac{1}{\tau_{inf}})$ | $\tau_{inf} = 5$ | 1/64 | Infected (pre-contagious) period (days) |

| 11 | $\theta_{con} = \exp(-\frac{1}{\tau_{con}})$ | $\tau_{con} = 3$ | 1/64 | Infectious (contagious) period (days) |
|---|---|---|---|---|
| **Clinical** | | | | |
| **12** | $\theta_{dev}$ | 1/3 | 1/64 | Pr(*symptoms* | *infected*) |
| **13** | $\theta_{sev}$ | 1/100 | 1/64 | Pr(*ARDS* | *symptomatic*) |
| **14** | $\theta_{sym} = \exp(-\frac{1}{\tau_{sym}})$ | $\tau_{sym} = 5$ | 1/64 | symptomatic period (days) |
| **15** | $\theta_{rds} = \exp(-\frac{1}{\tau_{rds}})$ | $\tau_{rds} = 12$ | 1/64 | acute RDS period (days) |
| **16** | $\theta_{fat}$ | 1/3 | 1/64 | Pr(*fatality* | *CCU*) |
| **17** | $\theta_{sur}$ | 1/16 | 1/64 | Pr(*survival* | *home*) |
| **Testing** | | | | |
| **18** | $\theta_{tft}$ | 500/100000 | 1/64 | Threshold: testing capacity (per capita) |
| **19** | $\theta_{sen}$ | 1/100 | 1/64 | Rate:    testing capacity (%) |
| **20** | $\theta_{del} = \exp(-\frac{1}{\tau_{del}})$ | $\tau_{del} = 2$ | 1/64 | Delay:    testing capacity (days) |
| **21** | $\theta_{tes}$ | 1/8 | 1/64 | Relative Pr(*tested* | *uninfected*) |

**Sources** (Huang et al., 2020; Mizumoto and Chowell, 2020; Russell et al., 2020; Verity et al., 2020; Wang et al., 2020a) and:

- https://www.statista.com/chart/21105/number-of-critical-care-beds-per-100000-inhabitants/
- https://www.gov.uk/guidance/coronavirus-covid-19-information-for-the-public
- http://www.imperial.ac.uk/mrc-global-infectious-disease-analysis/covid-19/

These prior expectations should be read as the effective rates and time constants as they manifest in a real-world setting. For example, a three-day period of contagion is shorter than the period that someone might be infectious (Wölfel et al., 2020)[12], on the (prior) assumption that they will self-isolate, when they realise they could be contagious.

## Initial conditions and population size

Further parameters are required to generate data, such as the size of the population and the number of people who are initially infected $(\theta_N, \theta_n)$ [13], which parameterise the initial state of the population (where $\otimes$ denotes

---

[12] Shedding of COVID-19 viral RNA from sputum can outlast the end of symptoms. Seroconversion occurs after 6-12 days but is not necessarily followed by a rapid decline of viral load.

[13] Table 1 also includes a parameter for the proportion of people who are initially immune, which we will call on later.

a Kronecker tensor product):

$$p_0^{loc} = [\tfrac{3}{4}, \tfrac{1}{4}, 0, 0]$$
$$p_0^{inf} \propto [10^6 \cdot \theta_N, \theta_n, 0, 0]$$
$$p_0^{clin} = [1, 0, 0, 0]$$
$$p_0^{test} = [1, 0, 0, 0]$$

$$p_0 = p_0^{loc} \otimes p_0^{inf} \otimes p_0^{clin} \otimes p_0^{test}$$

(1.10)

These parameters are unknown quantities that have to be estimated from the data; however, we still have to specify their prior densities. This begs the question: what kind of population are we trying to model? There are several choices here; ranging from detailed grid models of the sort used in weather forecasting (Palmer and Laure, 2013) and epidemiologic models (Ferguson et al., 2006). One could use models based upon partial differential equations; i.e., (Markov random) field models (Deco et al., 2008). In this technical report, we will choose a simpler option that treats a pandemic as a set of linked point processes that can be modelled as rare events. In other words, we will focus on modelling a single outbreak in a region or city and treat the response of the 'next city' as a discrete process *post hoc*. This simplifies the generative model; in the sense we only have to worry about the ensemble dynamics of the population that comprises one city. A complimentary perspective on this choice is that we are trying to model the first wave of an epidemic as it plays out in the first city to be affected. Any second wave can then be treated as the first wave of another city or region.

Under this choice, the population size can be set, *a priori*, to 1,000,000; noting that a small city comprises (by definition) a hundred thousand people, while a large city can exceed 10 million. Note that this is a prior expectation, the *effective* population size is estimated from the data: the assumption that the effective population size reflects the total population of a country is a hypothesis (that we will test later).

## The likelihood or observation model

The outcomes considered in Figure 2 are new cases (of positive tests and deaths) per day. These can be generated by multiplying the appropriate probability by the (effective) population size. The appropriate probabilities here are just the expected occupancy of positive test and deceased states, respectively. Because we are dealing with large populations, the likelihood of any observed daily count has a binomial distribution that can be approximated by a Gaussian density[14].

---

[14] This likelihood model can be finessed using a negative binomial distribution: MRC Centre for Global Infectious Disease Analysis: Report 13 (http://www.imperial.ac.uk/mrc-global-infectious-disease-analysis/covid-19/). However, a binomial is sufficient for our purposes.

$$o_t \sim B(n, \pi_t) \approx N(n\pi_t, n\pi_t(1 - \pi_t)) \Rightarrow$$
$$O_t = \sqrt{o_t} \sim N(\sqrt{n\pi_t}, I)$$

(1.11)

Here, outcomes are counts of rare events with a small probability $\pi \ll 1$ of occurring in a large population of size $n \gg 1$. For example, the likelihood of observing a timeseries of daily deaths can be expressed as a function of the model parameters as follows:

$$P(O \mid \vartheta) = P(O_0, \ldots, O_T \mid \vartheta) = \prod_0^T N(\sqrt{n\pi_t}, I)$$
$$n\pi_t = (10^6 \cdot \theta_N)(p_{deceased,t}^{clin} - p_{deceased,t-1}^{clin})$$
$$p_{t+1} = T(\vartheta, p_t)p_t$$

(1.12)

The advantage of this limiting (large population) case is that a (variance stabilising) square root transform of the data counts renders their variance unity. With the priors and likelihood model in place, we now have a full joint probability over causes (parameters) and consequences (outcomes). This is the generative model $P(O, \vartheta) = P(O \mid \vartheta)P(\vartheta)$. One can now use standard variational techniques (Friston et al., 2007) to estimate the posterior over model parameters and evaluate a variational bound on the model evidence or marginal likelihood. Mathematically, this is expressed as follows:

$$P(\vartheta) = N(\eta, C)$$
$$Q(\vartheta) = N(\mu, \Sigma)$$
$$Q(\vartheta) = \arg\max_Q F[Q, O]$$

(1.13)

$$F = \overbrace{E_Q[\ln P(O \mid \vartheta)]}^{accuracy} - \overbrace{D_{KL}[Q(\vartheta) \parallel P(\vartheta)]}^{complexity}$$
$$= E_Q[\ln P(\vartheta \mid O) + \ln P(O) - \ln Q(\vartheta)]$$
$$= \underbrace{\ln P(O)}_{evidence} - \underbrace{D_{KL}[Q(\vartheta) \parallel P(\vartheta \mid O)]}_{bound} \leq \underbrace{\ln P(O)}_{evidence}$$

These expressions show that maximising the variational free energy $F$ with respect to an approximate posterior $Q(\vartheta)$ renders the Kullback-Leibler (KL) divergence between the true and approximate posterior as small as possible. At the same time, the free energy becomes a lower bound on the log evidence. The free energy can then be used to compare different models, where any differences correspond to a log Bayes factor or odds ratio (Kass and Raftery, 1995; Winn and Bishop, 2005).

## Bayesian model comparison

One may be asking why we have chosen this particular state space and this parameterisation? Are there alternative model structures or parameterisations that would be more fit for purpose? The answer is that there will always be a better model, where 'better' is a model that has more evidence. This means that the model has to be optimised in relation to empirical data. This process is known as *Bayesian model comparison* based upon model evidence (Winn and Bishop, 2005). For example, in the above model we assumed that social distancing increases as a function of the proportion of the population who are infected (1.1). This stands in for a multifactorial influence on social behaviour that may be mediated in many ways. For example, government advice, personal choices, availability of transport, media reports of 'panic buying' and so on. So, what licenses us to model the causes of social distancing in terms of a probability that any member of the population is infected? The answer rests upon Bayesian model comparison. When inverting the model using data from countries with more than 16 deaths (see Figure 2), we obtained a log evidence (i.e., variational free energy) of -15701 natural units (nats). When replacing the cause of social distancing with the probability of encountering someone with symptoms—or the number of people testing positive— the model evidence fell substantially to -15969 and -15909 nats, respectively. In other words, there was overwhelming evidence in favour of infection rates as a primary drive for social distancing, over and above alternative models. We will return to the use of Bayesian model comparison later, when asking what factors determine differences between each country's response to the pandemic.

## Summary

Table 1 lists all the model parameters; henceforth, *DCM parameters*. In total, there are 21 DCM parameters. This may seem like a large number to estimate from the limited amount of data available (see Figure 2). The degree to which a parameter is informed by the data depends upon how changes in the parameter are expressed in data space. For example, increasing the effective population size will uniformly elevate the expected cases per day. Conversely, decreasing the number of initially infected people will delay the curve by shifting it in time. In short, a parameter can be identified if it has a relatively unique expression in the data. This speaks to an important point, the information in the data is not just in the total count—it is in the shape or form of the transient[15].

On this view, there are many degrees of freedom in a timeseries that can be leveraged to identify a highly parameterised model. The issue of whether the model is over parameterised or under parameterised is exactly the issue resolved by Bayesian model comparison; namely, the removal of redundant parameters to suppress model complexity and ensure generalisation: see (1.13)[16]. One therefore requires the best measures of model evidence. This is the primary motivation for using variational Bayes; here, variational Laplace (Friston et al., 2007). The variational free energy, in most circumstances, provides a better approximation than alternatives such as the widely used Akaike information criteria and the widely used Bayesian information criteria (Penny, 2012).

---

[15] A transient here refers to a transient perturbation to a system, characterising a response that evolves over time.

[16] Intuitively, this can be likened to a bat inverting its generative model of the world using the transients created by echo location. The shape of the transient contains an enormous amount of information, provided the bat has a good model of how echoes are generated. Exactly the same principle applies here: if one can find the right model, one can go beyond the immediate information in the data to make some precise inferences—based upon prior beliefs that constitute the generative model. This kind of abductive inference speaks to the importance of having a good forward or generative model—and the ability to select the best model based upon model evidence.

One special aspect of the model above is that it has absorbing states. For example, whenever one enters the morgue, becomes immune, dies or has a definitive test result, one stays in that state: see Figure 1. This is important, because it means the long-term behaviour of the model has a fixed point. In other words, we know what the final outcomes will be. These outcomes are known as endemic equilibria. This means that the only uncertainty is about the trajectory from the present point in time to the distant future. We will see later that—when quantified in terms of Bayesian credible intervals—this uncertainty starts to decrease as we go into the distant future. This should be contrasted with alternative models that do not parameterise the influences that generate outcomes and therefore call upon exogenous inputs (e.g., statutory changes in policy or changes in people's behaviour). If these interventions are unknown, they will accumulate uncertainty over time. By design, we elude this problem by including everything that matters within the model and parameterising strategic responses (like social distancing) as an integral part of the transition probabilities.

We have made the simplifying assumption that every country reporting new cases is, effectively, reporting the first wave of an affected region or city. Clearly, some countries could suffer simultaneous outbreaks in multiple cities. This is accommodated by an effective population size that could be greater than the prior expectation of 1 million. This is an example of finding a simple model that best predicts outcomes—that may not be a veridical reflection of how those outcomes were actually generated. In other words, we will assume that each country behaves *as if* it has a single large city of at-risk denizens. In the next section, we look at the parameter estimates that obtain by pooling information from all countries, with a focus on between country differences, before turning to the epidemiology of a single country (the United Kingdom).

Hitherto, we have focused on a generative model for a single city. However, in a pandemic, many cities will be affected. This calls for a hierarchical generative model that considers the response of each city at the first level and a global response at the second. This is an important consideration because it means, from a Bayesian perspective, knowing what happens elsewhere places constraints (i.e., Bayesian shrinkage priors) on estimates of what is happening in a particular city. Clearly, this rests upon the extent to which certain model parameters are conserved from one city to another—and which are idiosyncratic or unique. This is a problem of hierarchical Bayesian modelling or parametric empirical Bayes (Friston et al., 2016; Kass and Steffey, 1989). In the illustrative examples below, we will adopt a second level model in which key (log) parameters are sampled from a Gaussian distribution with a global (worldwide) mean and variance. From the perspective of the generative model, this means that to generate a pandemic, one first samples city-specific parameters from a global distribution, adds a random effect, and uses the ensuing parameters to generate a timeseries for each city.

# Parametric empirical Bayes and hierarchical models

This section considers the modelling of country-specific parameters, under a simple (general linear) model of between-country effects. This (second level) model requires us to specify which parameters are shared in a meaningful way between countries and which are unique to each country. Technically, this can be cast as the difference between *random* and *fixed effects*. Designating a particular parameter as a random effect means that this parameter was generated by sampling from a countrywide distribution, while a fixed effect is unique to each country. Under a general linear model, the distribution for random effects is Gaussian. In other words, to generate the parameter for a particular country, we take the global expectation and add a random Gaussian variate, whose variance has to be estimated under suitable hyperpriors. Furthermore, one has to specify systematic differences between countries in terms of independent variables; for example, does the latitude of a country have any systematic effect on the size of the at-risk population? The general linear model used here comprises a constant (i.e., the expectation or mean of each parameter over countries), the (logarithms of) total population size, and a series of independent variables based upon a discrete sine transform of latitude and longitude. The latter variables stand in for any systematic and geopolitical differences among countries that vary smoothly with their location. Notice that the total population size may or may not provide useful constraints on the effective size of the population at the first level. Under this hierarchical model, a bigger country may have a transport and communication infrastructure that could reduce the effective (at risk) population size. A hint that this may be the case is implicit in Figure 2, where there is no apparent relationship between the early incidence of deaths and total population size.

In the examples below, we treated the number of initial cases and the parameters pertaining to testing as fixed effects and all remaining parameters as random effects. The number of initial infected people determines the time at which a particular country evinces its outbreak. Although this clearly depends upon geography and other factors, there is no *a priori* reason to assume a random variation about an average onset time. Similarly, we assume that each country's capacity for testing was a fixed effect; thereby accommodating non-systematic testing or reporting strategies[17]. Note that in this kind of modelling, outcomes such as new cases can only be interpreted in relation to the probability of being tested and the availability of tests[18].

---

[17] This reflects concerns that data from different countries may not have been acquired or reported using the same criteria.

[18] These are two of several factors that conspire to produce the actual outcomes. The purpose of having a generative model is that reports of new cases may provide useful constraints on the latent causes of death rates. This means it is imperative to model the latent causes that interact in generating one sort of outcome, so that it can inform the causes of another. This fusion of different sorts of outcomes is the *raison d'être* for a generative model, when inferring their common underlying causes.

With this model in place, we can now use standard procedures for parametric empirical Bayesian modelling (Friston et al., 2016; Kass and Steffey, 1989) to estimate the second level parameters that couple between-country independent variables to country-specific parameters of the DCM. However, there are a large number of these parameters—that may or may not contribute to model evidence. In other words, we need some way of removing redundant parameters based upon Bayesian model comparison. This calls upon another standard procedure called *Bayesian model reduction* (Friston et al., 2018; Friston et al., 2016). In brief, Bayesian model reduction allows one to evaluate the evidence for a model that one would have obtained if the model had been reduced by removing one or more parameters. The key aspect of Bayesian model reduction is that this evidence can be evaluated using the posteriors and priors of a parent model that includes all possible parameters. There are clearly an enormous number of combinations of parameters that one could consider. Fortunately, these can be scored quickly and efficiently using Bayesian model reduction, by making use of Savage-Dickey density ratios (Friston and Penny, 2011; Savage, 1954). Because Bayesian model reduction scores the effect of changing the precision of priors—on model evidence—it can be regarded as an automatic Bayesian sensitivity analysis, also known as robust Bayesian analysis (Berger, 2011).

Figure 3 shows the results of this analysis. The upper panels show the posterior probability of 256 models that had the greatest evidence (shown as a log posterior in the upper left panel). Each of these models corresponds to a particular combination of parameters that have been 'switched off', by shrinking their prior variance to zero. By averaging the posterior estimates in proportion to the evidence for each model, —known as *Bayesian model averaging* (Hoeting et al., 1999)—we can eliminate redundant parameters and thereby provide a simpler explanation for differences among countries. This is illustrated in the lower panels, which show the posterior densities before (left) and after (right) Bayesian model reduction. These estimates are shown in terms of their expectation or *maximum a posteriori* (MAP) value (as blue bars), with 90% Bayesian credible intervals (as pink bars).

The first 21 parameters are the global expectations of the DCM parameters. The remaining parameters are the coefficients that link various independent variables at the second level to the parameters of the transition probabilities at the first. Note that a substantial number of second level parameters have been removed; however, many are retained. This suggests that there are systematic variations over countries in certain random effects at the country level. Figure 4 provides an example based upon the largest effect mediated by the independent variables. In this analysis, latitude (i.e., distance from the South Pole) appears to reduce the effective size of an at-risk population. In other words, countries in the northern hemisphere have a smaller effective population size, relative to countries in the southern hemisphere. Clearly, there may be many reasons for this; for example, systematic differences in temperature or demographics.

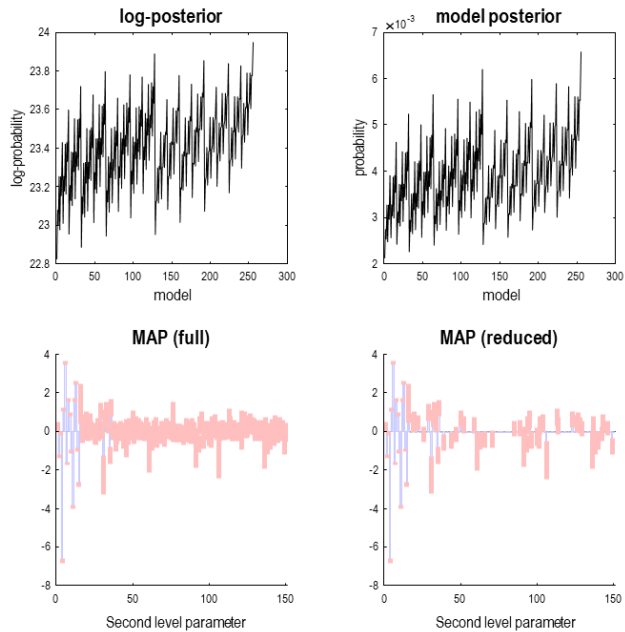

**Figure 3: Bayesian model reduction.** This figure reports the results of Bayesian model reduction. In this instance, the models compared are at the second or between-country level. In other words, the models compared contained all combinations of (second level) parameters (a parameter is removed by setting its prior variance to zero). If the model evidence increases—in virtue of reducing model complexity—then this parameter is redundant. The upper panels show the relative evidence of the most likely 256 models, in terms of log evidence (left panel) and the corresponding posterior probability (right panel). Redundant parameters are illustrated in the lower panels by comparing the posterior expectations before and after the Bayesian model reduction. The blue bars correspond to posterior expectations, while the pink bars denote 90% Bayesian credible intervals. The key thing to take from this analysis is that a large number of second level parameters have been eliminated. These second level parameters encode the effects of population size and geographical location, on each of the parameters of the generative model. The next figure illustrates the nonredundant effects that can be inferred with almost 100% posterior confidence.

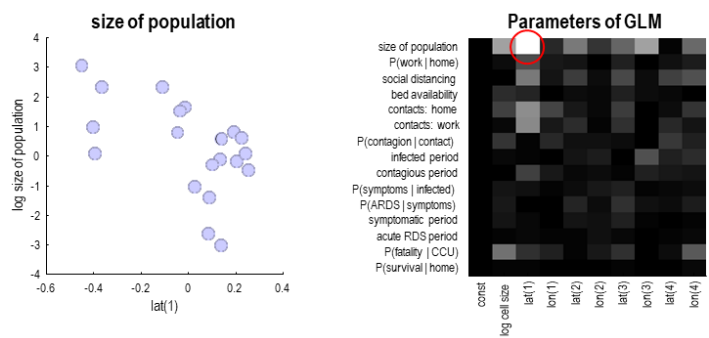

**Figure 4: between country effects**. This figure shows the relationship between parameters of the generative model and the explanatory variables in a general linear model of between country effects. The left panel shows a regression of country-specific DCM parameters on the independent variable that had the greatest absolute value; namely, the contribution of an explanatory variable to a model parameter. Here, the effective size of the population appears to depend upon the latitude of a country. The right panel shows the absolute values of the GLM parameters in matrix form, showing that the effective size of the population was most predictable (the largest values are in white), though not necessarily predictable by total population size. The red circle highlights the parameter mediating the relationship illustrated in the left panel.

Figure 5 shows the Bayesian parameter averages (Litvak et al., 2015) of the DCM parameters over countries. The posterior density (blue bars and pink lines) are supplemented with the prior expectations (red bars) for comparison. The upper panel shows the MAP estimates of log parameters, while the lower panel shows the same results in terms of scale parameters. The key thing to take from this analysis is the tight credible intervals on the parameters, when averaging in this way. According to this analysis, the number of effective contacts at home is about three people, while this increases by an order of magnitude to about 30 people when leaving home. The symptomatic and acute respiratory distress periods have been estimated here at about five and 13 days respectively, with a delay in testing of about two days. These are the values that provide the simplest explanation for the global data at hand—and are in line with empirical estimates[19].

---

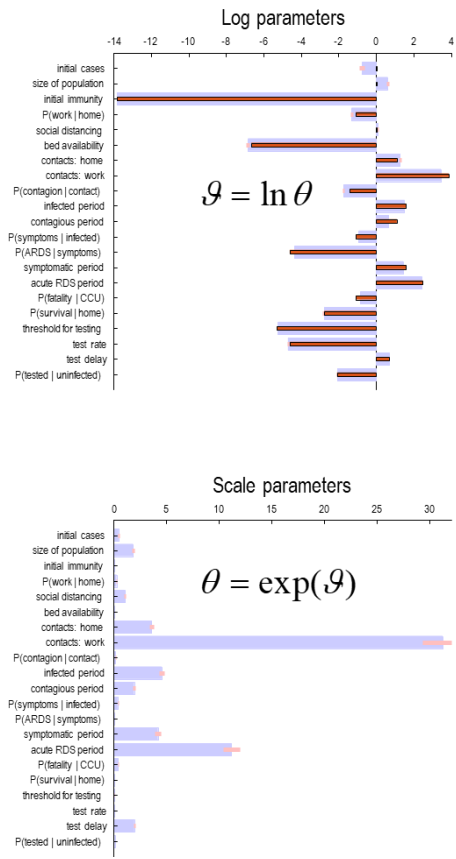

**Figure 5: Bayesian parameter averages**. This figure reports the Bayesian parameter averages over countries following a hierarchical or parametric empirical Bayesian analysis that tests for—and applies shrinkage priors to—posterior parameter estimates for each country. The upper panel shows the parameters as estimated in log space, while the lower panel shows the same results for the corresponding scale (nonnegative) parameters. The blue bars report posterior expectations, while the thinner red bars in the upper panel are prior expectations. The pink bars denote 90% Bayesian credible intervals. One can interpret these parameters as the average value for any given parameter of the generative model, to which a random (country-specific) effect is added to generate the ensemble dynamics for each country. In turn, these ensemble distributions determine the likelihood of various outcome measures under large number (i.e., Gaussian) assumptions.

Figure 6 shows the country-specific parameter estimates for 12 of the 21 DCM parameters. These posterior densities were evaluated under the empirical priors from the parametric empirical Bayesian analysis above. As one might expect—in virtue of the second level effects that survived Bayesian model reduction—there are some substantial differences between countries in certain parameters. For example, the effective population size in the United States of America is substantially greater than elsewhere at about 25 million (the population in New York state is about 19.4 million). The effective population size in the UK (dominated by cases in London) is estimated to be about 2.5 million (London has a population of about 7.5 million)[20]. Social distancing seems to be effective and sensitive to infection rates in France but much less so in Canada. The efficacy of social distancing in terms of the difference between the number of contacts at home and work is notably attenuated in the United Kingdom—that has the greatest number of home contacts and the least number of work contacts. Other notable differences are the increased probability of fatality in critical care evident in China. This is despite the effective population size being only about 2.5 million. Again, these assertions are not about actual states of affairs. These are the best explanations for the data under the simplest model of how those data were caused.

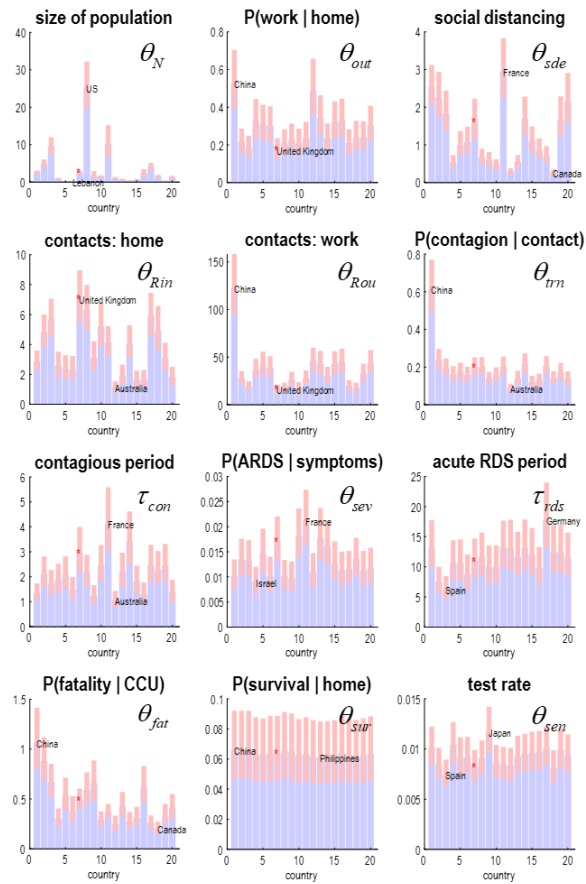

**Figure 6: differences among countries**. This figure reports the differences among countries in terms of selected parameters of the generative model, ranging from the effective population size, through to the probability of testing

its denizens. The blue bars represent the posterior expectations, while the pink bars are 90% Bayesian credible intervals. Notice that these intervals are not symmetrical about the mean because we are reporting scale parameters—as opposed to log parameters. For each parameter, the countries showing the smallest and largest values are labelled. The red asterisk denotes the country considered in the next section (the United Kingdom). The next figure illustrates the projections, in terms of new deaths and cases, based upon these parameter estimates. The order of the countries is listed in Figure 2.

## Summary

This level of modelling is important because it enables the data or information from one country to inform estimates of the first level (DCM) parameters that underwrite the epidemic in another country[21]. This is another expression of the importance of having a hierarchical generative model for making sense of the data. Here, the generative model has latent causes that span different countries, thereby enabling the fusion of multimodal data from multiple countries (e.g., new test or death rates). Two natural questions now arise. Are there any systematic differences between countries in the parameters that shape epidemiological dynamics—and what do these dynamics or trajectories look like?

This concludes our brief treatment of between country effects, in which we have considered the potentially important role of Bayesian model reduction in identifying systematic variations in the evolution of an epidemic from country to country. The next section turns to the use of hierarchically informed estimates of DCM parameters to characterise an outbreak in a single country.

# Dynamic causal modelling of a single country

This section drills down on the likely course of the epidemic in the UK, based upon the posterior density over DCM parameters afforded by the hierarchical (parametric empirical) Bayesian analysis of the previous section (listed in Table 2). Figure 7 shows the expected trajectory of death rates, new cases, and occupancy of CCU beds over a six-month (180 day) period. These (posterior predictive) densities are shown in terms of an expected trajectory and 90% credible intervals (blue line and shaded areas, respectively). The black dots represent empirical data (available at the time of writing). Notice that the generative model can produce outcomes that may or may not be measured. Here, the estimates are based upon the new cases and deaths in Figure 2.

The panels on the left show that our confidence about the causes of new cases is relatively high during the period for which we have data and then becomes uncertain in the future. This reflects the fact that the data are informing those parameters that shaped the initial transient, whereas other parameters responsible for the late peak and subsequent trajectory are less informed. Notice that the uncertainty about cumulative deaths itself accumulates. On this analysis, we can be 90% confident that in five weeks, between 13,000 and 22,000 people may have died. Relative to the total population, the proportion of people dying is very small; however, the cumulative death rates in absolute numbers are substantial, in relation to seasonal influenza (indicated with broken red lines). Although cumulative death rates are small, they are concentrated within a short period of time, with near-identical CCU needs—with the risk of over-whelming available capacity (not to mention downstream effects from blocking other hospital admissions to prioritise the pandemic).

---

[21] Or, indeed, a previous pandemic, such as the 2009 H1H1 pandemic. We will return to this in the conclusion.

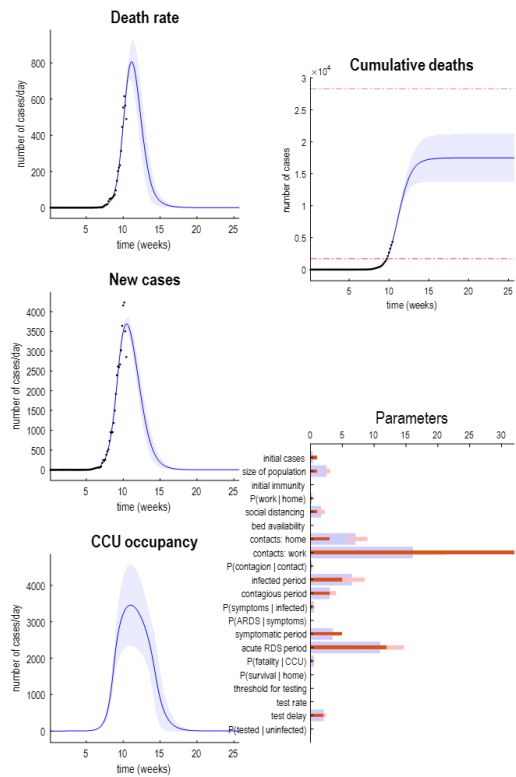

**FIGURE 7**

**Figure 7: projected outcomes**. This figure reports predicted[22] new deaths and cases (and CCU occupancy) for an exemplar country; here, the United Kingdom. The panels on the left show the predicted outcomes as a function of

---

[22] We will use predictions—as opposed to projections—when appropriate, to emphasise the point that the generative model is not a timeseries model, in the sense that the unknown quantities (DCM parameters) do not change with time. This means the there is uncertainty about predictions in the future *and the past*, given uncertainty about the parameters (see Figure 7). This should be contrasted with the notion of forecasting or projection; however, predictions in the future, in this setting, can be construed as projections.

weeks. The blue lines correspond to the expected trajectory, while the shaded areas are 90% Bayesian credible intervals. The black dots represent empirical data, upon which the parameter estimates are based. The lower right panel shows the parameter estimates for the country in question. As in previous figures, the prior expectations are shown as pink bars over the posterior expectations (and credible intervals). The upper right panel illustrates the equivalent expectations in terms of cumulative deaths. The dotted red lines indicate the number of people who died from seasonal influenza in recent years[23]. The key point to take from this figure is the quantification of uncertainty inherent in the credible intervals. In other words, uncertainty about the parameters propagates through to uncertainty in predicted outcomes. This uncertainty changes over time because of the nonlinear relationship between model parameters and ensemble dynamics. By model design, one can be certain about the final states; however, uncertainty about cumulative death rates itself accumulates. The mapping from parameters, through ensemble dynamics to outcomes is mediated by latent or hidden states. The trajectory of these states is illustrated in the next figure.

The underlying latent causes of these trajectories are shown in Figure 8. The upper panels reproduce the expected trajectories of the previous figure, while the lower panels show the underlying latent states in terms of expected rates or probabilities. For example, the social distancing measures are expressed in terms of an increasing probability of being at home, given the accumulation of infected cases in the population. During the peak expression of death rates, the proportion of people who are immune (herd immunity) increases to about 30% and then asymptotes at about 90%. This period is associated with a marked increase in the probability of developing symptoms (peaking at about 11 weeks, after the first reported cases). Interestingly, under these projections, the number of people expected to be in critical care should not exceed capacity: at its peak, the upper bound of the 90% credible interval for CCU occupancy is approximately 4200, this is within the current CCU capacity of London (corresponding to the projected capacity of the temporary Nightingale Hospital[24] in London, UK).

---

[23] Public Health England estimates that on average 17,000 people have died from the flu in England annually between 2014/15 and 2018/19. However, yearly deaths vary widely, from a high of 28,330 in 2014/15 to a low of 1,692 in 2018/19 (broken red lines in Figure 7).

[24] Note, only 2800 beds are ventilator/ITU beds.

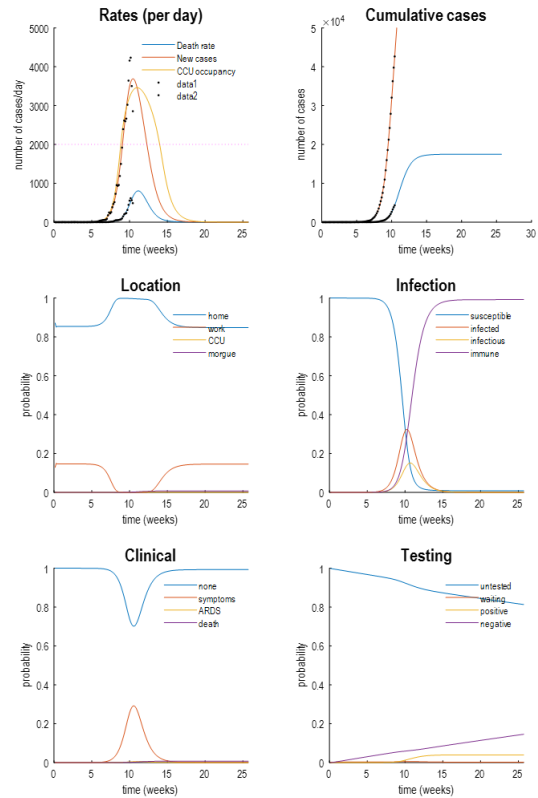

**Figure 8: latent causes of observed consequences**. The upper panels reproduce the expected trajectories of the previous figure, for an example country (here the United Kingdom). The expected death rate is shown in blue, new cases in red, predicted recovery rate in orange and CCU occupancy in yellow. The black dots correspond to empirical data. The lower four panels show the evolution of latent (ensemble) dynamics, in terms of the expected probability of being in various states. The first (location) panel shows that after about 5 to 6 weeks, there is sufficient evidence for the onset of an episode to induce social distancing, such that the probability of being found at work falls, over a couple of weeks to negligible levels. At this time, the number of infected people increases (to about 32%) with a concomitant probability of being infectious a few days later. During this time, the probability of becoming immune increases monotonically and saturates at about 20 weeks. Clinically, the probability of becoming symptomatic rises to about 30%, with a small probability of developing acute respiratory distress and, possibly death (these probabilities are very small and cannot be seen in this graph). In terms of testing, there is a progressive increase in the number of people tested, with a concomitant decrease in those untested or waiting for their results. Interestingly, initially the number of negative tests increases monotonically, while the proportion of positive tests starts to catch up during the peak of the episode. Under these parameters, the entire episode lasts for about 10 weeks, or less than three months. The broken red line in the upper left panel shows the typical number of CCU beds available to a well-resourced city, prior to the outbreak.

| parameter | Mean | Units | Upper | Lower |
|---|---|---|---|---|
| initial cases | 0.33 | | 0.19 | 0.57 |
| size of population | 2.49 | M | 1.99 | 3.11 |
| Pr(work \| home) | 17 | % | 12 | 23 |
| social distancing | 1.60 | | 1.159 | 2.2 |
| contacts: home | 7.01 | | 5.49 | 8.95 |
| contacts: work | 16.02 | | 12.12 | 21.19 |
| Pr(contagion \| contact) | 19 | % | 15 | 25 |
| infected period | 6.44 | Days | 4.87 | 8.51 |
| contagious period | 2.93 | Days | 2.15 | 3.99 |
| Pr(symptoms \| infected) | 47 | % | 34 | 65 |
| Pr(ARDS \| symptoms) | 1.70 | % | 1.31 | 2.20 |
| symptomatic period | 3.45 | Days | 2.55 | 4.68 |
| acute RDS period | 10.89 | Days | 8.07 | 14.70 |
| Pr(fatality \| CCU) | 48 | % | 39 | 60 |
| Pr(survival \| home) | 6.36 | % | 4.56 | 8.87 |
| test delay | 2.02 | Days | 1.668 | 2.45 |
| Pr(tested \| uninfected) | 0.12 | % | 10 | 15 |

It is natural to ask which DCM parameters contributed the most to the trajectories in Figure 8. This is addressed using a *sensitivity analysis*. Intuitively, this involves changing a particular parameter and seeing how much it affects the outcomes of interest. Figure 9 reports a sensitivity analysis of the parameters in terms of their direct contribution to cumulative deaths (upper panel) and how they interact (lower panel). These are effectively the gradient and Hessian matrix (respectively) of predicted cumulative deaths. The bars in the upper panel pointing to the left indicate parameters that decrease total deaths. These include social distancing and bed availability, which are—to some extent—under our control. Other factors that improve fatality rates include the symptomatic and acute respiratory distress periods and the probability of surviving outside critical care. These, at the present time, are not so amenable to intervention. Note that initial immunity has no effect in this analysis because we clamped the initial values to zero with very precise priors. We will relax this later. First, we look at the effect of social distancing by simulating the ensemble dynamics under increasing levels of the social distancing exponent (i.e., the sensitivity of our social distancing and self-isolation behaviour to the prevalence of the virus in the community).

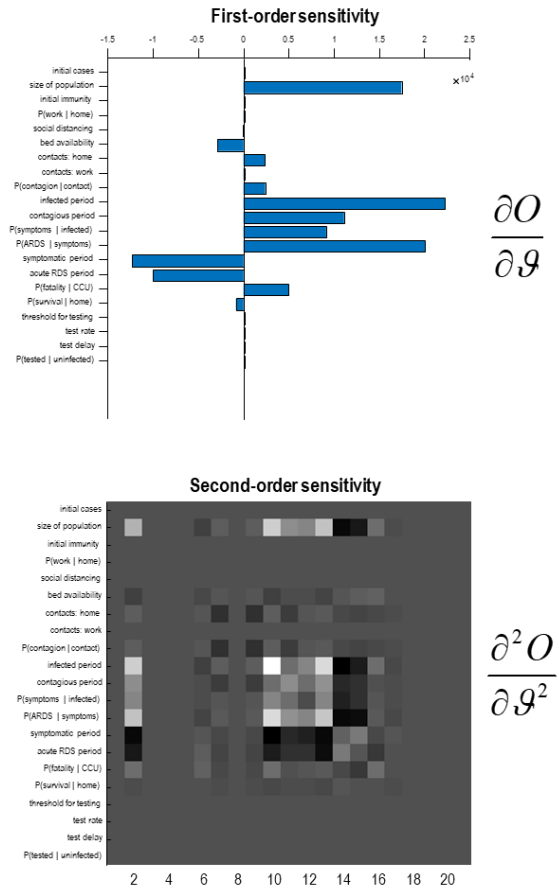

**Figure 9: sensitivity analysis**. These panels show the change in outcome measures—here cumulative deaths—with respect to model parameters (upper panel: first order derivatives. lower panel: second order derivatives). The bar charts in the upper panel are the derivatives of outcomes with respect to each of the parameters. Positive values (on the right) exacerbate new cases when increased, while negative values (on the left) decrease new cases. As one might expect, increasing social distancing, bed availability and the probability of survival outside critical care, tend to decrease death rate. Interestingly, increasing both the period of symptoms and ARDS decreases overall death rate, because (in this compartmental model) keeping someone alive for longer in a CCU reduces fatality rates (as long as capacity is not exceeded). The lower panel shows the second order derivatives. These reflect the effect of one parameter on the effect of another parameter on total deaths. For example, the effects of bed availability and fatality in CCU are positive, meaning that the beneficial (negative) effects of increasing bed availability—on total deaths—decrease with fatality rates.

It may be surprising to see that social distancing has such a small effect on total deaths (see upper panel in Figure 9). However, the contribution of social distancing is in the context of how the epidemic elicits other responses; for example, increases in critical care capacity. Quantitatively speaking, increasing social distancing only delays the expression of morbidity in the population: it does not, in and of itself, decrease the cumulative cost (although it buys time to develop capacity, treatments, and primary interventions). This is especially the case if there is no effective limit on critical care capacity, because everybody who needs a bed can be accommodated. This speaks to the interaction between different causes or parameters in generating outcomes. In the particular case of the UK, the results in Figure 4 suggest that although social distancing is in play, self-isolation appears limited. This is because the number of contacts at home is relatively high (at over five); thereby attenuating the effect of social distancing. In other words, slowing the spread of the virus depends upon reducing the number of contacts by social distancing. However, this will only work if there is a notable difference between the number of contacts at home and work. One can illustrate this by simulating the effects of social distancing, when it makes a difference.

Figure 10 reproduces the results in Figure 8 but for 16 different levels of the social distancing parameter, while using the posterior expectation for contacts at home (of about four) from the Bayesian parameter average. Social distancing is expressed in terms of the probability of being found at home or work (see the panel labelled *location*). As we increase social distancing the probability and duration of being at home during the outbreak increases. This flattens the curve of death rates per day from about 600 to a peak of about 400. This is the basis of the mitigation ('curve flattening') strategies that have been adopted worldwide. The effect of this strategy is to reduce cumulative deaths and prevent finite resources being overwhelmed. In this example, from about 17,000 to 14,000, potentially saving about 3000 people. This is roughly four times the number of people who die in the equivalent period due to road traffic accidents. Interestingly, these (posterior predictive) projections suggest that social distancing can lead to an endgame in which not everybody has to be immune (see the middle panel labelled *infection*). We now look at herd immunity using the same analysis.

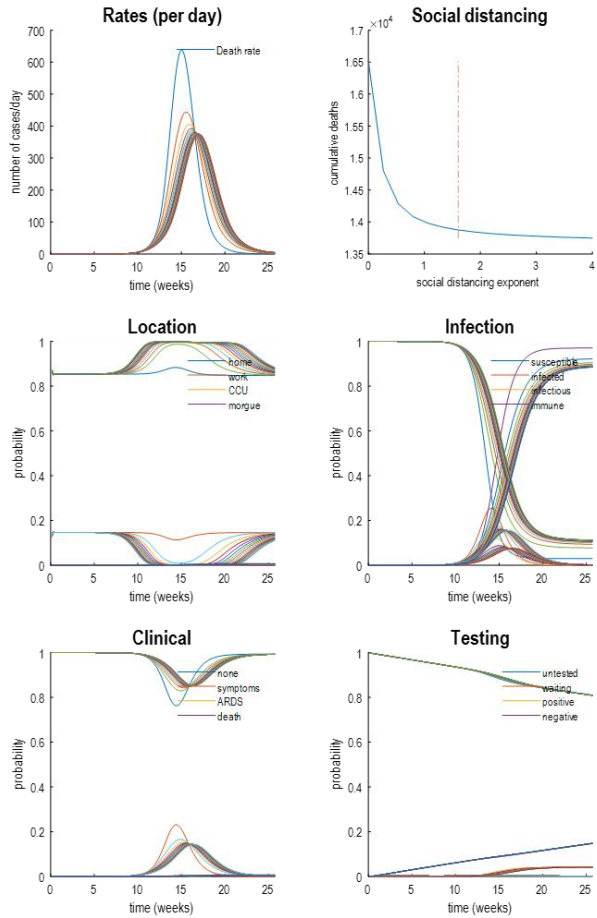

**Figure 10: the effects of social distancing**. This figure uses the same format as Figure 9. However, here trajectories are reproduced under different levels of social distancing; from zero through to four (in 16 steps). This parameter is the exponent applied to the probability of not being infected. In other words, it scores the sensitivity of social distancing to the prevalence of the virus in the population. In this example (based upon posterior expectations for the United Kingdom and Bayesian parameter averages over countries), death rates (per day) decrease progressively with social distancing. The cumulative death rate is shown as a function of social distancing in the upper right panel. The vertical line corresponds to the posterior expectation of the social distancing exponent for this country. These results suggest that social distancing relieves pressure on critical care capacities and ameliorates cumulative deaths by about 3000 people. Note that these projections are based upon an effective social distancing policy at home, with about four contacts. In the next figure, we repeat this analysis but looking at the effect of herd immunity.

Figure 11 reproduces the results in Figure 10 using the United Kingdom posterior estimates – but varying the initial (herd) immunity over 16 levels from, effectively, 0 to 100%. The effects of herd immunity are marked, with cumulative deaths ranging from about 18,000 with no immunity to very small numbers with a herd immunity of about 70%. The broken red lines in the upper right panel are the number of people dying from seasonal influenza (as in Figure 7). These projections suggest that there is a critical level of herd immunity that will effectively avert an epidemic; in virtue of reducing infection rates, such that the spread of the virus decays exponentially. If we now return to Figure 8, it can be seen that the critical level of herd immunity will, on the basis of these projections, be reached 2 to 3 weeks after the peak in death rates. At this point—according to the model—social distancing starts to decline as revealed by an increase in the probability of being at work. We will put some dates on this trajectory by expressing it as a narrative in the conclusion.

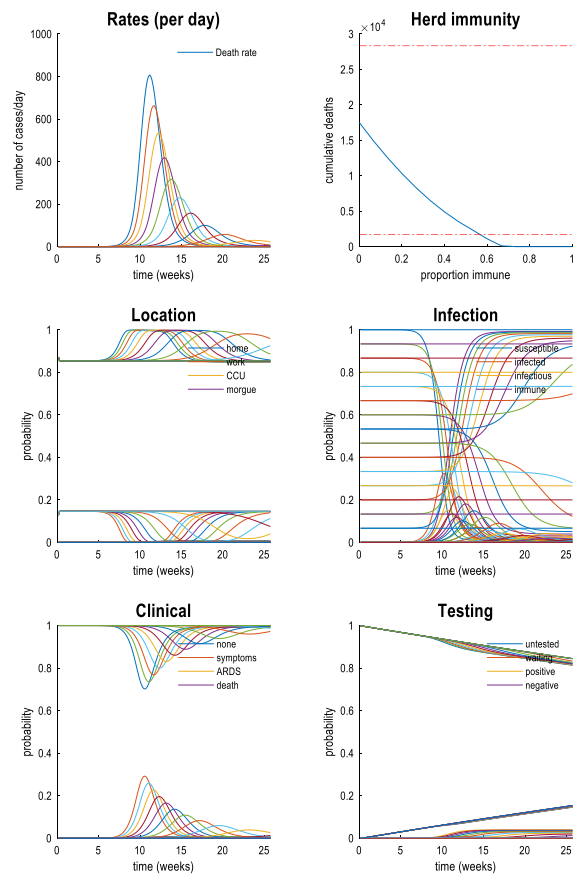

**Figure 11: herd immunity**. This figure reproduces the format of the previous figure. However, here, we increased

the initial proportion of the at-risk population who were initially immune. Increasing the initial immunity dramatically decreases death rates with a fall in the cumulative deaths from several thousand to negligible levels with an initial herd immunity of about 70%. The dashed lines in the upper panel shows the equivalent deaths over the same time period due to seasonal flu (based upon 2014/2014 and 2018/2019 figures). The lower deaths due to seasonal flu would require an initial herd immunity of about 60%.

From a modelling perspective, the influence of initial herd immunity is important because it could form the basis of modelling the spread of the virus from city to another—and back again. In other words, more sophisticated generative models can be envisaged, in which an infected person from one city is transported to another city with a small probability or rate. Reciprocal exchange between cities, (and ensuing 'second waves') will then depend sensitively on the respective herd immunities in different regions. Anecdotally, other major pandemics, without social isolation strategies, have almost invariably been followed by a second peak that is as high (e.g., the 2009 H1N1 pandemic), or higher, than the first. Under the current model, this would be handled in terms of a second region being infected by the first city and so on; like a chain of dominos or the spread of a bushfire (Rhodes and Anderson, 1998; Zhang and Tang, 2016). Crucially, the effect of the second city (i.e., wave) on the first will be sensitive to the herd immunity established by the first wave. In this sense, it is interesting to know how initial levels of immunity shape a regional outbreak, under idealised assumptions.

Figure 12 illustrates the interaction between immunity and viral spread as characterised by the effective reproduction rate, $R$ (a.k.a. number or ratio); see (1.9). This figure plots the predicted death rates for the United Kingdom and the accompanying fluctuations in $R$ and herd immunity, where both are treated as outcomes of the generative model. The key thing to observe is that with low levels of immunity, $R$ is fairly high at around 2.5 (current estimates of the basic reproduction ratio[25] $R_0$, in the literature, range from 1.4 to 3.9). As soon as social distancing comes into play, $R$ falls dramatically to almost 0. However, when social distancing is relaxed some weeks later, $R$ remains low due to the partial acquisition of herd immunity, during peak of the epidemic. Note that herd immunity in this setting pertains to, and only to, the effective or at-risk population: 80% herd immunity a few months from onset would otherwise be overly optimistic, compared to other *de novo* pandemics; e.g., (Donaldson et al., 2009). On the other hand, an occult herd immunity (i.e. not accompanied by symptoms) is consistent with undocumented infection and rapid dissemination (Li et al., 2020). Note that this way of characterising the spread of a virus depends upon many variables (in this model, two factors and three parameters). And can vary from country to country. Repeating the above analysis for China gives a much higher an initial or basic reproduction rate, which is consistent with empirical reports (Steven et al., 2020).

---

[25] The basic reproduction ratio is a constant that scores the spread of a contagion in a susceptible population. This corresponds to the effective reproduction ratio at the beginning of the outbreak, when everybody is susceptible. See Figure 12.

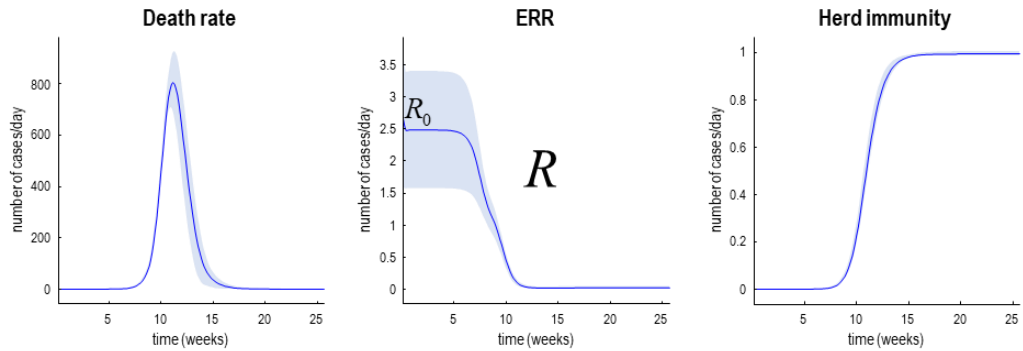

**Figure 12: effective reproduction ratio**. This figure plots the predicted death rates for the United Kingdom from Figure 6 and the concomitant fluctuations in the effective reproduction rate (*R*) and herd immunity. The blue lines represent the posterior expectations while the shaded areas correspond to 90% credible intervals.

This concludes our characterisation of projections for what is likely to happen and what could happen under different scenarios for a particular country. In the final section, we revisit the confidence with which these posterior predictive projections can be made.

# Predictive Validity

Variational approaches—of the sort described in this technical report—use all the data at hand to furnish statistically efficient estimates of model parameters and evidence. This contrasts with alternative approaches based on cross-validation. In the cross-validation schemes, model evidence is approximated by cross-validation accuracy. In other words, the evidence for a model is scored by the log likelihood that some withheld or test data can be explained by the model. Although model comparison based upon a variational evidence bound renders cross-validation unnecessary, one can apply the same procedures to demonstrate predictive validity. Figure 13 illustrates this by fitting partial timeseries from one country (Italy) using the empirical priors afforded by the parametric empirical Bayesian analysis. These partial data comprise the early phase of new cases. If the model has predictive validity, the ensuing posterior predictive density should contain the data that was withheld during estimation. Figure 13 presents an example of forward prediction over a 10-day period that contains the peak death rate. In this example, the withheld data are largely within the 90% credible intervals, speaking to the predictive validity of the generative model. There are two caveats here: first, similar analyses using very early timeseries from Italy failed to predict the peak, because of insufficient (initial) constraints in the data. Second, the credible intervals probably suffer from the well-known overconfidence problem in variational Bayes, and the implicit mean field approximation

(MacKay, 2003)[26].

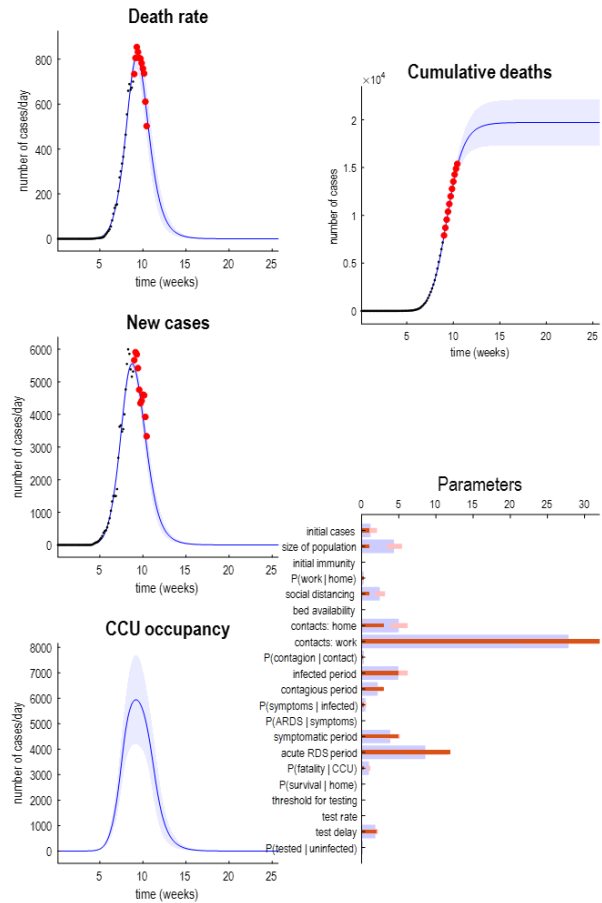

**Figure 13: predictive validity**. This figure uses the same format as Figure 7; however, here, the posterior estimates are based upon partial data, from early in the timeseries for an exemplar country (Italy). These estimates were obtained under (parametric) empirical Bayesian priors. The red dots show outcomes that were not used to estimate the expected trajectories (and credible intervals). This example illustrates the predictive validity of the estimates for a 10-day period following the last datapoint, which capture the rise to the peak of new cases.

---

[26] Note further that the credible intervals can include negative values. This is an artefact of the way in which the intervals are computed: here, we used a first-order Taylor expansion to propagate uncertainty about the parameters through to uncertainty about the outcomes. However, because this generative model is non-linear in the parameters, high-order terms are necessarily neglected.

# Conclusions

We have rehearsed variational procedures for the inversion of a generative model of a viral epidemic—and have extended this model using hierarchical Bayesian inference (parametric empirical Bayes) to deal with the differential responses of each country, in the context of a worldwide pandemic. The utility of such modelling is self-evident: one can predict, with a quantified degree of confidence, what may happen in the near future. For example, under the posterior beliefs based upon data at the time of writing, one can sketch a narrative for what people in London may experience over the forthcoming weeks. This narrative would be something like the following:

*"Based on current data, reports of new cases in London are expected to peak on April 5, followed by a peak in death rates around April 10 (Good Friday). At this time, critical care unit occupancy should peak, approaching—but not exceeding—capacity, based on current predictions and resource availability. At the peak of death rates, the proportion of people infected (in London) is expected to be about 32%, which should then be surpassed by the proportion of people who are immune at this time. Improvements should be seen by May 8, shortly after the May bank holiday, when social distancing will be relaxed. At this time herd immunity should have risen to about 80%, about 12% of London's population will have been tested. Just under half of those tested will be positive. By June 12, death rates should have fallen to low levels with over 90% of people being immune and social distancing will no longer be a feature of daily life."*

Clearly, this narrative is entirely conditioned on the generative model used to make these predictions (e.g., the assumption of lasting immunity, which may or may not be true). The narrative is offered in a deliberately definitive fashion to illustrate the effect of resolving uncertainty about what will happen. It has been argued that many deleterious effects of the pandemic are mediated by uncertainty. This is probably true at both a psychological level—in terms of stress and anxiety (Davidson, 1999; McEwen, 2000; Peters et al., 2017)—and at an economic level in terms of 'loss of confidence' and 'uncertainty about markets'. Put simply, the harmful effects of the coronavirus pandemic are not just what will happen but the effects of the uncertainty about what will happen. This is a key motivation behind procedures that quantify uncertainty, above and beyond being able to evaluate the evidence for different hypotheses about what will happen.

One aspect of this is reflected in rhetoric such as "there is no clear exit strategy". It is reassuring to note that, if one subscribes to the above model, there is a clear exit strategy inherent in the self-organised mitigation[27] afforded by herd immunity. For example, within a week of the peak death rate, there should be sufficient herd immunity to preclude any resurgence of infections in, say, London. The term 'self-organised' is used carefully here. This is because we are part of this process, through the effect of social distancing on our location, contact with infected people and subsequent dissemination of COVID-19. In other words, this formulation does not preclude strategic (e.g., nonpharmacological) interventions; rather, it embraces them as part of the self-organising ensemble dynamics.

---

[27] This technical report does not distinguish between *mitigation* and *suppression*: in the generative model under consideration, both go hand-in-hand.

# Outstanding issues

There are several outstanding issues that present themselves:

The generative model—at both the first and second level—needs to be explored more thoroughly. At the first level, this may entail the addition of other factors; for example, splitting the population into age groups or different classes of clinical vulnerability. Procedurally, this should be fairly simple, by specifying the DCM parameters for each age group (or cohort) separately and precluding transitions between age groups (or cohorts). One could also consider the fine graining of states within each factor. For example, making a more careful distinction between being in and not in critical care (e.g., being in self-isolation, being in a hospital, community care home, rural or urban location and so on). At the between city or country level, the parameters of the general linear model could be easily extended to include a host of demographic and geographic independent variables. Finally, it would be fairly straightforward to use increasingly fine-grained outcomes, using regional timeseries, as opposed to country timeseries (these data are currently available from: https://github.com/CSSEGISandData/COVID-19).

Another plausible extension to the hierarchical model is to include previous outbreaks of MERS and SARS (Middle East and Severe Acute Respiratory Syndrome, respectively) in the model. This would entail supplementing the timeseries with historical (i.e., legacy) data and replicating the general linear model for each type of virus. In effect, this would place empirical priors or constraints on any parameter that shares characteristics with MERS-CoV and SARS-CoV.

In terms of the model parameters—as opposed to model structure—more precise knowledge about the underlying causes of an epidemic will afford more precise posteriors. In other words, more information about the DCM parameters can be installed through adjusting the prior expectations and variances. The utility of these adjustments would then be assessed in terms of model evidence. This may be particularly relevant as reliable data about bed occupancy, proportion of people recovered, *etc* becomes available.

A key aspect of the generative model used in this technical report is that it precludes any exogenous interventions of a strategic sort. In other words, the things that matter are built into the model and estimated as latent causes. However, prior knowledge about fluctuating factors, such as closing schools or limiting international air flights, could be entered by conditioning the DCM parameters on exogenous inputs. This would explicitly install intervention policies into the model. Again, these conditions would only be licensed by an increase in model evidence (i.e., through comparing the evidence for models with and without some structured intervention). This may be especially important when it comes to modelling future interventions, for example, a 'sawtooth' social distancing protocol. A simple example of this kind of extension would be including a time dependent increase in the capacity for testing: at present, constraints on testing rates are assumed to be constant.

A complementary approach would be to explore models in which social distancing depends upon variables that can be measured or inferred reliably (e.g., the rate of increase of people testing positive) and optimise the parameters of the ensuing model to minimise cumulative deaths. In principle, this should provide an operational equation that could be regarded as an adaptive (social distancing) policy, which accommodates as much as can be inferred about the epidemiology as possible.

A key outstanding issue is the modelling of how one region (or city) affects another—and how the outbreak spreads from region to region. This may be an important aspect of these kinds of models; especially when it comes to modelling second waves as 'echoes' of infection, which are reflected back to the original epicentre. As noted above, the ability of these echoes to engender a second wave may be sensitively dependent on the herd immunity established during the first episode. Herd immunity is therefore an important (currently latent or unobserved) state. This speaks to the importance of antibody testing in furnishing empirical constraints on herd immunity. In turn, this motivates antibody testing, *even if the specificity and sensitivity of available tests are low*. Sensitivity and specificity are not only part of generative models, they can be estimated along with the other model parameters. In this setting, the role of antibody testing would be to provide data for population modelling and strategic advice—not to establish whether any particular person is immune or not (e.g., to allow them to go back to work).

Finally, it would be useful to assess the construct validity of the variational scheme adopted in dynamic causal modelling, in relation to schemes that do not make mean field approximations. These schemes usually rely upon some form of sampling (e.g., Markov Chain Monte Carlo sampling) and cross-validation. Cross-validation accuracy can be regarded as a useful but computationally expensive proxy for model evidence and is the usual way that modellers perform automatic Bayesian computation. Given the prevalence of these sampling based (non-variational) schemes, it would be encouraging if both approaches converged on roughly the same predictions. The aim of this technical report is to place variational schemes on the table, so that construct validation becomes a possibility in the short-term future.

## Software note

The figures in this technical report can be reproduced using annotated (MATLAB) code that is available as part of the free and open source academic software SPM (https://www.fil.ion.ucl.ac.uk/spm/). This software package has a relatively high degree of validation; being used for the past 25 years by over 5000 scientists in the neurosciences. The routines are called by a demonstration script that can be invoked by typing >> DEM_COVID. At the time of writing, these routines are undergoing software validation in our internal source version control system - that will be released in the next public release of SPM (and via GitHub at https://github.com/spm/).

# Acknowledgements

This work was undertaken by members of the Wellcome Centre for Human Neuroimaging, UCL Queen Square Institute of Neurology. The Wellcome Centre for Human Neuroimaging is supported by core funding from Wellcome [203147/Z/16/Z]. A.R. is funded by the Australian Research Council (Refs: DE170100128 and DP200100757). A.J.B. is supported by a Wellcome Trust grant WT091681MA. CL is supported by an MRC Clinician Scientist award (MR/R006504/1).

The authors declared no conflicts of interest.